\documentclass[aps,pre,showpacs,amsmath,amsfonts,amssymb,superscriptaddress,preprint]{revtex4}

\usepackage[english]{babel}
\usepackage{graphicx}

\usepackage{amsthm}
\usepackage{amssymb}
\usepackage{SIunits}
\usepackage{amsmath,mathrsfs}
\usepackage{hhline}
\usepackage{fancyhdr}

\usepackage{appendix}

\everymath{\displaystyle}
\everymath{\displaystyle}

\newcommand{\D}{\mathrm{d}}
\newcommand{\affA}{Aix Marseille Univ, CNRS, CPT, Marseille, France}
\newcommand{\affB}{CNRS Centre de Physique Th\'eorique UMR7332,
13288 Marseille, France}

\newcommand{\affG}{CNR - Consiglio Nazionale delle Ricerche - Istituto dei Sistemi Complessi, 50019, Sesto Fiorentino, Italy}
\newcommand{\affH}{Institut f\"{u}r Theoretische Physik, Technische Universit\"{a}t Berlin, Hardenbergstr. 36, 10623 Berlin,
Germany}

\begin{document}

\title{Collective behavior of oscillating electric dipoles}

\author{Simona Olmi}
\email{simona.olmi@fi.isc.cnr.it}
\affiliation{\affH}\affiliation{\affG}
\author{Matteo Gori}
\email{gori@cpt.univ-mrs.fr}
\affiliation{\affA}\affiliation{\affB}
\author{Irene Donato}
\email{irene.donato@cpt.univ-mrs.fr} 
\affiliation{\affB}\
\author{Marco Pettini}
\email{pettini@cpt.univ-mrs.fr}
\affiliation{\affA}\affiliation{\affB}

\begin{abstract}
The present work reports about the dynamics of a collection of randomly distributed, and randomly oriented, 
oscillators in 3D space, coupled by an interaction potential falling as $1/r^3$, where $r$ stands for the inter-particle 
distance. This model schematically represents a collection of identical biomolecules, coherently vibrating at some 
common frequency, coupled with a $-1/r^3$ potential stemming from the electrodynamic interaction between 
oscillating dipoles. The oscillating dipole moment of each molecule being a direct consequence of its coherent 
(collective) vibration.
By changing the average distance among the molecules, neat and substantial changes in the power spectrum 
of the time variation of a collective observable are found. As the average intermolecular distance can be varied 
by changing the concentration of the solvated molecules, and as the collective variable investigated is proportional
to the projection of the total dipole moment of the model biomolecules on a coordinate plane, we have found a 
prospective experimental strategy of spectroscopic kind to check whether the mentioned intermolecular electrodynamic 
interactions can be strong enough to be detectable, and thus to be of possible relevance to biology.
\end{abstract}
\date{\today}
\maketitle

%PART I
\section{INTRODUCTION}
Let us quickly summarize what motivates the present work. The starting point is the observation of the enormous efficiency, rapidity, and robustness against environmental disturbances, 
of the complex network of biochemical reactions in living cells. At the same time it is hardly conceivable that this pattern of interactions/reactions is
driven and regulated only by random encounters between cognate partners \cite{stroppolo}. In fact, on the basis of several estimates \cite{northrup,schreiber}, 
in many cases the high efficiency that biomolecules display when moving toward their specific targets and sites of action can hardly be the result of thermal fluctuations (Brownian motion) alone: 
biochemical players "need to know" where to go and when. Therefore, in order to accelerate these encounters, selective forces acting at a long distance ("long" means possibly up to 
some hundreds of nanometers) are needed. In the physico-chemical conditions typical of the cytoplasm (large value of the static dielectric constant of water, strong Debye shielding 
due to high concentrations of freely moving ions) electrostatic forces are ruled out; to the contrary, electrodynamic interactions of sufficiently high frequency can be effective. 
Quite a long time ago, it was surmised \cite{froehlich} that if each of the cognate partners of a biochemical reaction would undergo a collective vibrational oscillation 
(involving all the atoms or a large fraction of them in each molecule) at the same or almost the same frequency, then the associated giant dipole vibrations could excite a sufficiently 
intense and resonant (thus selective) electrodynamic attractive interaction \cite{pre3}. This would be the basic mechanism of molecular  recruitment at a distance, beyond all the well-known 
short-range forces (chemical, covalent bonding, H-bonding, Van der Waals). Unfortunately, because of technological limitations, an experimental proof or refutation of this possibility has 
been for a long time and is still sorely lacking. These long range electrodynamic interactions are predicted by standard classical electrodynamics, thus they necessarily exist, 
the point is whether these can attain a sufficient strength to overcome all the dissipation mechanisms that would be activated together with the collective vibration \cite{pre3}. 
In our preliminary investigations in \cite{pre1} and \cite{pre2, Nardecchia2017} we have put forward the idea that an answer to this conundrum could come from the study of how the diffusion behavior 
of biomolecules in solution could change when their concentration is varied (that is, when the average intermolecular distance is varied) as a consequence of the action of surmised electrodynamic interactions.
The experimental technique envisaged in \cite{pre2, Nardecchia2017} was Fluorescence Correlation Spectroscopy (FCS), a well established experimental technique \cite{Webb1974,Schwille2008,Elson2011}. 
In the present paper we report about a possible alternative/complementary viable experimental procedure for an assessment of the strength - thus of the potential biological relevance - 
of resonant electrodynamic intermolecular interactions. The paper is organized as follows: in Section II the model is defined and discussed, while in Sec. III  we report the outcomes of 
the Molecular Dynamics simulations of the chosen model and we comment on the observed phenomenology. Section IV is devoted to some concluding remarks  about the results presented throughout the present paper. 

\section{The model}

\subsection{Model for the biomolecule}

This work aims to study the emergence of collective phenomena in a system
of mutually interacting classical electric dipole oscillators out-of-thermal equilibrium. 
This is intended to be a little step further in the same direction
of \cite{pre3} where the hypothesis have been explored of the possibility that
long-range classical electrodynamic interactions can take place in living
matter at molecular level.\\
As in that case, an oversimplified model for biomolecules has
been used, where the level of accuracy taken into account is suited to
this feasibility study.
Each biomolecule has been described as an oscillating electric
dipole composed of two material points, each of them with a mass
$m$ and the same absolute value $Ze$ of the electric charge but
with opposite sign. The position of the positive and negative
charged particles of the i-th biomolecule are respectively
$\mathbf{r}_{+,i}$ and $\mathbf{r}_{-,i}$.
The position of its center of mass of each biomolecule
is indicated by $\mathbf{R}_{i}$ while the direction of each dipole is
\begin{equation}
\widehat{\mathbf{r}}_{i}=\dfrac{\mathbf{r}_{+,i}-\mathbf{r}_{-,i}}{\|\mathbf{r}_{+,i}-\mathbf{r}_{-,i}\|};
\end{equation}
both have been considered to be fixed, so that the 
charged particle of each biomolecule are constrained to oscillate
along their joining line.\\
Both the constraints on $\mathbf{R}_{i}$ and $\widehat{\mathbf{r}}_{i}$
seem to be quite strong with respect to a realistic biological molecular
system where particles both diffuse (time dependence $\mathbf{R}_{i}$) and
rotate due to the collisions with the surrounding water molecules.
These assumptions are justified by the comparison of characteristic
time scales for collective giant dipole oscillations of a single biomolecule
with characteristic time scales given by the translational and rotational diffusion
(See Appendix \ref{sec:TimScales} for more details).
It follows that the only dynamical variable is the mutual distance
$r_{i}=(\mathbf{r}_{+,i}-\mathbf{r}_{-,i})\cdot \widehat{\mathbf{r}}_{i}$
between the two centres of charge. The electric dipole moment is given by
$\mathbf{p}_{i}(t)=Ze r_{i}(t)\widehat{\mathbf{r}}_i$.
Despite its simplicity, this model takes into account the main features
we are interested in: collective effects on the dynamics of giant
dipole oscillations emerging in a system of biomolecules coupled through
electrodynamic long-range interactions.\\
For each dipole representing a biomolecule, an effective potential 
$V(r_{i})=V_{eff}(\|\mathbf{r}_{+,i}-\mathbf{r}_{-,i}\|)$ is supposed to be exerted among material charged points. A stable equilibrium configuration
is supposed to be attained for $r_{i}=r_{i,eq}$ such that the effective potential
takes the form
\begin{equation}
\label{eq:EffPot_bio}
V_{eff}(r_{i})\approx\dfrac{1}{2}m\omega_{i}^2 (r_{i}-r_{i,0})^2+\dfrac{1}{4}m\dfrac{\omega_i^2}{\Lambda^2}(r_{i}-r_{i,0})^4,
\end{equation}
where the parameter $\Lambda$ is the characteristic length of the oscillation
amplitude for the emergence of non-harmonic contributions.
So the effective potential of \eqref{eq:EffPot_bio} takes in account both
harmonic and non-harmonic contributions in the oscillation of the electric
dipole. The non-harmonic contribution has been included for two main
reasons: firstly, it accounts for the exchange of energy of the
main collective mode with other vibrational normal modes of the
biomolecule; secondly, it has been included in order to prevent
instability of the oscillations when the electric dipoles, representing
biomolecules, are strongly coupled among them.

\subsection{Mutual quasi-electrostatic interactions among biomolecules}

The physical picture behind the model we intend to analise is
an ensemble of oscillating biomolecules in watery solutions
in presence of freely moving ions.
Since the declared interest of this work is to study collective
phenomena mediated by long-range interactions among biomolecules,
we neglect any electrostatic effect due to Debye screening
effects. We can easily make this assumpion as the electrostatic field is exponentially suppressed on
a length scale of the order of some Angstroms in real biological
systems. It follows that, for the intermolecular average distance
range we are interested in ($\sim 10^2-10^3\AA$), the contribution
of electrostatic field is negligible.
On the contrary electrodynamics fields are not screened in watery
systems in presence of freely moving ions, as it can be inferred both
from theoretical works and from dielectric spectroscopic experiments for
sufficiently high frequencies ($\omega> 10^2 MHz$).
As mentioned before the expected frequency for the collective
oscillation of a biomolecule is around $0.1-1 THz$, thus largely
above the upper frequency threshold for important
screening effects on electrodynamic fields.
Collective phenomena are more probably expected in systems
of resonant oscillators: for such a reason, a system of $N$
identical biomolecules (oscillators) has been considered.
Moreover, resonance of electric dipole oscillators, describing
biomolecules, has been argued to be a  necessary condition in order 
to activate long range dipole-dipole ($\sim R_{ij}^{-3}$) electrodynamic 
interactions \cite{pre3}.\\
In our very simple model the force acting on 
each charge barycentre of the $i$-th electric dipole
due to the $j$-th dipole is given by
\begin{equation}
\mathbf{F}_{CED}(\mathbf{r}_{\pm,i};\mathbf{R}_{j})=Ze\mathbf{E}_{CED}(\mathbf{r}_{\pm,i};\mathbf{r}_{j})\,.
\end{equation}
where $\mathbf{E}_{CED}(\mathbf{r};\mathbf{R}_{j})$ is the value of the electric field in $\mathbf{r}$
generated by the $j$-th dipole whose center is in $\mathbf{R}_j$.
According to the Classical Electrodynamics (CED), if we assume valid the dipole approximation,
i.e. $\|\mathbf{r}-\mathbf{R}_j\|\gg r_{j}$, the expression for the electric field takes the form
\begin{equation}
\label{eq:E_CED_jthdipole_exact}
\begin{split}
&\mathbf{E}_{CED}(\mathbf{r};\mathbf{R}_{j})=\displaystyle{\int_{0}^{+\infty}\,
	\D\omega\,\,\dfrac{\exp\left[ i \omega \left(t\pm\sqrt{\epsilon(\omega)}\|\mathbf{r}-\mathbf{R}_{j}\|/c\right)\right]}{4\pi\epsilon(\omega)	\|\mathbf{r}-\mathbf{R}_{j}\|^3}}\\
&\times\Biggr\{\left[3\widehat{\mathbf{n}}_{j}(\mathbf{r})(\mathbf{p}_{j}(\omega)\cdot \widehat{\mathbf{n}}_ {j}(\mathbf{r}))-\mathbf{p}_{j}(\omega)\right]\left(1\mp\dfrac{i\omega\sqrt{\epsilon(\omega)}\|\mathbf{r}- \mathbf{R}_{j}\|}{c}\right)+\\
&-\left[\mathbf{p}_{j}(\omega)-\widehat{\mathbf{n}}_{j}(\mathbf{r})(\mathbf{p}_{j}(\omega)\cdot\widehat{\mathbf{n}}_
{j}(\mathbf{r}))\right]\dfrac{\omega^2 \epsilon(\omega)\| \mathbf{r}-\mathbf{R}_{j} \|^2}{c^2}\Biggr\}\,.
\end{split}
\end{equation}
where $c$ is the speed of light, $\widehat{\mathbf{n}}_{j}=\mathbf{r}-
\mathbf{R}_j/(\|\mathbf{r}-\mathbf{R}_j\|)$ is direction
joining the center of dipole $\mathbf{R_j}$ to $r$, $\mathbf{p}_{j}(\omega)$
is the Fourier Transform  of the electric dipole moment of the $j$-th biomolecule in time domain and
$\epsilon(\omega)$ is the dielectric constant of the medium.\\
For the range of frequencies we explore ($\omega\sim\Omega\approx 1 THz$), the dielectric constant
of an electrolytic aqueous solution can assumed to be real
$\mathfrak{Re}\left(\epsilon(\omega)\right)\gg\mathfrak{Im}\left(\epsilon(\omega)\right)$ 
and approximatively constant $\epsilon_{WS}(\Omega)\approx 3$. 
Moreover both the intermolecular average distance $R_{ij}\approx 10^3 \angstrom$
and the characteristic linear dimensions $r_{0}\approx 10\angstrom$ are much 
smaller than the characteristic wavelength of the electromagnetic field
$\lambda=  2\pi c/(\epsilon \omega)\simeq 5\times 10^{7}\angstrom$.
This allows to assume that the electromagnetic field has the same value
for both centers of charge of each biomolecule, i.e.
$\mathbf{E}_{CED}(\mathbf{r}_{+,i};\mathbf{R}_{j})=\mathbf{E}_{CED}(\mathbf{r}_{-,i};\mathbf{r}_{j})=\mathbf{E}_{CED}(\mathbf{R}_{i};\mathbf{R}_{j})$, 
and that any retardation effect can be neglected, i.e. $R_{ij}/\lambda\ll 1 $.
With these approximations the acceleration of the $i$-th dipole is directed along $\widehat{\mathbf{r}}_{i}$
and due to the interaction with the $j$-th dipole reads as
\begin{equation}
\begin{split}
&\left(m \dfrac{\mathrm{d}^2 r_{i}}{\mathrm{d}t^2}\right)_{CED}=\left(m \dfrac{\mathrm{d}^2 \mathbf{r}_{+,i}}{\mathrm{d}t^2}-m\dfrac{\mathrm{d}^2 \mathbf{r}_{-,i}}{\mathrm{d}t^2}\right)_{CED}\cdot\widehat{\mathbf{r}}_{i}=2Ze\sum_{j\neq i}\mathbf{E}_{CED}(\mathbf{R}_{i};\mathbf{R}_{j})\cdot\widehat{\mathbf{r}}_{i}=\\
&=2 Ze\sum_{j\neq i}\int_{0}^{+\infty}\,\D\omega\,\,\dfrac{\exp\left(i \omega t\right)}{4\pi\epsilon_{WS} R_{ij}^3}\left[3(\widehat{\mathbf{n}}_{ji}\cdot \widehat{\mathbf{r}}_{i})(\mathbf{p}_{j}(\omega)\cdot \widehat{\mathbf{n}}_ {ji}(\mathbf{r}))-\mathbf{p}_{j}(\omega)\cdot \widehat{\mathbf{r}}_{i} \right]=\\
&=2 (Ze)^2\sum_{j\neq i}\int_{0}^{+\infty}\,\D\omega\,\,\dfrac{\exp\left(i \omega t\right)}{4\pi\epsilon_{WS} R_{ij}^3}\left[3(\widehat{\mathbf{n}}_{ji}\cdot \widehat{\mathbf{r}}_{i})(\widehat{\mathbf{r}}_{j}\cdot \widehat{\mathbf{n}}_ {ji})-(\widehat{\mathbf{r}}_{j}\cdot \widehat{\mathbf{r}}_{i}) \right]r_{j}(\omega)=\\
&=2 (Ze)^2\sum_{j\neq i}\dfrac{\left[3(\widehat{\mathbf{n}}_{ji}\cdot \widehat{\mathbf{r}}_{i})(\widehat{\mathbf{r}}_{j}\cdot \widehat{\mathbf{n}}_ {ji})-(\widehat{\mathbf{r}}_{j}\cdot \widehat{\mathbf{r}}_{i}) \right]}{4\pi\epsilon_{WS} R_{ij}^3}r_{j}(t)
=\sum_{j\neq i}m\omega^{2}_{ij}\zeta_{ij}r_{j}(t),
\end{split}
\end{equation}
where $\widehat{\mathbf{n}}_{ji}=\dfrac{\mathbf{R}_{j}-\mathbf{R}_{i}}{R_{ij}}$ is the direction joining
the electric dipoles,
\begin{equation}
\omega^{2}_{ij}= \displaystyle{\frac{2 Z_{i}Z_{j}e^{2}}{4\pi\epsilon_{\mathrm{WS}}m R_{ij}^3}}
\end{equation}
is a characteristic frequency describing the strength of the dipole-dipole interactions,
\begin{equation}
\zeta_{ij}=\left[3(\widehat{\mathbf{n}}_{ji}\cdot \widehat{\mathbf{r}}_{i})(\widehat{\mathbf{r}}_{j}\cdot \widehat{\mathbf{n}}_ {ji})-(\widehat{\mathbf{r}}_{j}\cdot \widehat{\mathbf{r}}_{i}) \right]
\end{equation}
is a geometrical factor depending of the orientation of the electric dipoles and $r_{j}(\omega)$ is the Fourier Transform of $r_{j}(t)$.

\section{Study of synchronization in presence of thermal bath and external source}
\subsection{Biological watery environment as thermal bath}
This work is inspired by the request for observables in real biological
systems at molecular level that can detect the presence of long-range 
electrodynamics interactions among biomolecules. As all biomolecules in
real biological environment are in watery solution, we have to take into
account the presence of surrounding water molecules.
Though recent studies reveal that the water in biological system can have a highly non trivial behaviour with
respect to electrodynamic fields generated by the electric dipole of
biomolecules \cite{DelGiudice1984,Meister2014,McDermott2017,montagner2017,Kurian2017}, in this article we will assume the surrounding water to play simply the role of a thermal bath.
As a consequence of this, the presence of water molecules can be schematized
via the introduction of a stochastic noise (thermal fluctuations) and a viscous friction term (dissipation)
in the equation of motion for oscillating electric dipoles.
In particular friction viscous forces are due to the aqueous surrounding medium
considered as a homogeneous fluid with viscosity $\eta_{w}$.
We assume that the expression of the viscous force is given by Stokes'
Law acting on each barycentre of electric charge (positive and negative)
\begin{equation}
\label{eq:viscforce_bary}
\mathbf{F}_{\mathrm{visc},i\pm}=-\gamma_{i} \frac{\D\mathrm{r}_{i,\pm}}{\D t} \qquad
\gamma_{i}=6\pi \eta_{W} \mathcal{R}_{i}
\end{equation}
where $\mathcal{R}_{i}$ is the hydrodynamic radius of a typical biomolecule ($\sim 10 \AA$). 
From eq.\eqref{eq:viscforce_bary} it follows that the acceleration
on the dipole length is given by
\begin{equation}
\left(m\dfrac{\mathrm{d}^2 r_i}{\mathrm{d}t^2}\right)_{FR}=\left(m\dfrac{\mathrm{d}^2}{\mathrm{d}t^2}
\left(\mathbf{r}_{i,+}-\mathbf{r}_{i,-}\right)\right)_{FR}\cdot\widehat{\mathbf{r}}_{i}=\left(\mathbf{F}_{\mathrm{visc},i+}-
\mathbf{F}_{\mathrm{visc},i-}\right)\cdot\widehat{\mathbf{r}}_{i}=-\gamma_{i} \dfrac{\D\mathrm{r}_{i}}{\D t}.
\end{equation}
On the other hand the stochastic forces are due to the collision of water molecules
and freely moving ions on the biomolecules and they correspond to the realization 
of a thermal bath at temperature $T$. In particular these forces, acting directly
on the charge barycentres of each biomolecules, can be described according to the following expression
\begin{equation}
\mathbf{F}_{\mathrm{stoch},i\pm}=\Xi_{i} \boldsymbol{\xi}_{i,\pm}(t) \qquad
\Xi=\sqrt{2 k_{B}T \gamma_{i}},
\end{equation}
where $\boldsymbol{\xi}_{i}(t)$ represents white noise whose characteristics along each Cartesian component $ \alpha,\beta=x,y,z$
are given by
\begin{equation}
\left\langle\left(\xi(t)_{i,\pm}\right)_{\alpha}\right\rangle_{t}=0 \qquad
\left\langle\left(\xi(t)_{i,\pm}\right)_{\alpha}\left(\xi(t')_{j,\pm}\right)_{\beta}\right\rangle_{t}=\delta(t-t')\delta_{ij}\delta_{\alpha\beta}\left(\delta_{++}+\delta_{--}-\delta_{+-}-\delta_{-+}\right)
\end{equation}
The minus sign in the correlation term is due to the constrains we impose for the noise
\begin{equation}
\boldsymbol{\xi}_{i,+}(t)=-\boldsymbol{\xi}_{i-}(t),
\end{equation}
constrains that allows to easily calculate the stochastic force along the dipole direction
\begin{equation}
\left(m\dfrac{\D^2 r_i }{\D t^2}\right)_{ST}=\left(\boldsymbol{\xi}_{i,+}(t)-\boldsymbol{\xi}_{i,-}(t)\right)\cdot \widehat{\mathbf{r}}_i=2 \boldsymbol{\xi}_{i,+}(t)\cdot\widehat{\mathbf{r}}_i=2\Xi_{i}\xi_{i}(t)\,.
\end{equation}

\subsection{Exteral forcing to produce out-of-thermal equilibrium conditions}
In \cite{pre3} it has been shown that long-range interactions among biomolecules can be exerted if the system of oscillating
dipoles is maintained in out-of-thermal equilibrium.
To achieve this goal a forcing term $F_{NE,i}(t)$ has been included in the equations of
motion for the electric dipoles in order to ensure an external injection of energy.
The explicit form of the force $F_{NE,i}(t)$ depends on the specific process that is chosen to inject energy into the system.
In particular, a possible mechanism that has been used recently in THz spectroscopy 
experiments to detect collective giant oscillations in biomolecules, is the injection of energy in vibrational
modes through the vibrational decay of the excited fluorochromes attached to each biomolecules \cite{NardecchiaTorres}.
This process can be represented choosing the following explicit form for the forcing term
\begin{equation}
F_{NE,i}(t)=A _{NE,i}\omega_{\mathrm{pul}}\,\, f_{\mathrm{pul}}(t;\omega_{\mathrm{pul}},\phi_i)
\end{equation}
where $f_{\mathrm{pul}}$ is a pulse-like function of the form
\begin{equation}
f_{\mathrm{pul}}(t;\omega_{\mathrm{pul}},\phi_i)=\frac{1}{2\pi}\sum_{i=1}^{n_{\mathrm{pul}}} a_{n} \left[1+\cos\left(\omega_{\mathrm{pul}}t+\phi_{i} \right)\right]^{n_{\mathrm{pul}}} 
\qquad a_{n}=\frac{2^n (n !)^2}{(2 n)!}.
\end{equation}
The coefficients in the former equation have been chosen such that the integral of the function $f_{\mathrm{pul}}$ over a period $T_{\mathrm{pul}}=2\pi\omega_{\mathrm{pul}}^{-1}$ 
respects the following normalization
\begin{equation}
\int_{0}^{\frac{2\pi}{\omega_{\mathrm{pul}}}} \,\, f_{\mathrm{pul}}(t;\omega_{\mathrm{pul}},\phi_i) \D t =\frac{1}{\omega_{\mathrm{pul}}} \,.
\end{equation} 
With this choice it is clear that $A_{NE,i}$ corresponds to the momentum transferred by the fluorochrome to the protein in a time $2\pi \omega_{\mathrm{pul}}^{-1}$.
The energy losses in vibrational decay can be estimated to be of the order $\Delta E_{\mathrm{pul}}=h\Delta \nu_{\mathrm{fluor}}$ where $\Delta \nu_{fluor}$ is the difference among frequencies 
of absorbed and emitted light by the flourochrome and $h$ is the Planck constant; consequently, if $m_{\mathrm{fluor}}$ is the mass of the fluorochrome, the momentum transferred to the biomolecule 
can be approximated by
\begin{equation}
\Delta (m_{i} \dot{r}_{i})\approx \sqrt{2h\Delta\nu m_{\mathrm{fluor}}}=A_{NE,i}=A_{NE}.
\end{equation}

\subsection{Equation of motion for the system of oscillating interacting dipoles}
The equations of motion that describe the dynamics of the system with mutually oscillating dipoles are
\begin{equation}
\label{eq:DipoleEqMotion}
\begin{split}
m\dfrac{\D^2 r_{i}}{\D t^2}=&-m\omega_{0}^2\left(r_{i}-r_{i0}\right)-m\dfrac{\omega_{0}^2}{\Lambda}\left(r_{i}-r_{i0}\right)^3+
\sum_{j\neq i}m\omega_{ij}^2\zeta_{ij}r_{j}+\\
&-\gamma \dfrac{\D r_{i}}{\D t}+2\Xi\xi(t)+F_{NE,i}(t) \qquad \qquad \forall i=1,...,N\\
\end{split}
\end{equation}
where all the biomolecules are assumed to be identical so that they all have the 
same characteristic frequencies $\omega_{i}=\omega_{0}$ and $\Lambda_{i}=\Lambda$.\\
In order to simplify the discussion we introduce the following scales
\begin{equation}
\label{eq:adimens_units}
m=\mu \widetilde{m}, \qquad t=\dfrac{\tau}{\omega_{0}} , \qquad r_{i}=\lambda x_{i} 
\end{equation}
that substituted in eq.\eqref{eq:DipoleEqMotion} yield to
\begin{equation}
\label{eq:Dipole_MotionEquation_Adim}
\begin{split}
 \frac{\D^2
x_{i}}{\D\tau^2}=&-\left(x_{i}-x_{i0}\right)-\dfrac{\left(x_{i}-x_{i0}\right)^3}{\widetilde{\Lambda}^2}-
\Omega_{\mathrm{frict},i}\dfrac{\D
x_{i}}{\D\tau}+\sum_{j\neq
i}^{N}\Omega^{2}_{ij}\zeta_{ij}x_{j}+\widetilde{\Psi}_{i}\widetilde{\xi}_{i}(t)+\\
&+\Omega_{\mathrm{pul}}\mathcal{A}_{NE}\,\,f_{\mathrm{pul}}(\tau ;\Omega_{\mathrm{pul}},\phi_i)
\qquad \forall i=1,...,N\\
\end{split}
\end{equation}
where 
\begin{equation}
\begin{split}
&\tilde{\Lambda}=\dfrac{\Lambda}{\lambda},\,\,\Omega_{ij}^2=\dfrac{\omega_{ij}^2}{\omega_0^2},\,\,\widetilde{\mathcal{R}}_i=\dfrac{\mathcal{R}_i}{\lambda},\,\,\widetilde{\eta}_W=\dfrac{\eta_W \lambda}{\mu \omega_0},\,\,\Omega_{\mathrm{frict},i}=\dfrac{6\pi\widetilde{\mathcal{R}}_i\widetilde{\eta}_W}{\tilde{m}_i},\,\, \mathcal{E}_{\mathrm{bath}}=\dfrac{k_{B}T}{\mu\lambda^2\omega_{0}^2},\\
&\widetilde{\xi}_{i}=\omega_{0}^{-1/2}\xi_i ,\,\,\, \tilde{\Psi}_i=\left(\dfrac{48\pi\mathcal{E}_{\mathrm{bath}}\widetilde{\mathcal{R}}_i\widetilde{\eta}_W}{\widetilde{m}_i^2}\right)^{1/2},\,\,
\Omega_{\mathrm{pul}}=\frac{\omega_{\mathrm{pul}}}{\omega_{0}},\,\, \mathcal{E}_{\mathrm{pul}}=\frac{h \Delta \nu_{\mathrm{fluorr}}}{\mu\omega_{0}^2\lambda^2},\\
&\widetilde{m}_{\mathrm{fluor}}=\dfrac{m_{\mathrm{fluor}}}{\mu},\,\, \mathcal{A}_{NE}=\left(\dfrac{\mathcal{E}_{\mathrm{pul}}\widetilde{m}_{fluor}}{\widetilde{m}_{i}^2}\right)^{1/2}.
\end{split}
\end{equation}

\subsection{Choice of numerical parameters in eq.\eqref{eq:Dipole_MotionEquation_Adim}}

The numerical values of parameters that appear in eq. \eqref{eq:Dipole_MotionEquation_Adim} 
have been estimated for a realistic biological system.
In particular the characteristic fundamental scales for the system have been fixed as following: i) the typical mass scale of a biomolecule
$\mu=1.66\times 10^{-24}\mathrm{Kg}=1\mathrm{KDa}$; ii) the characteristic length scale of a biomolecule
$\lambda=10^{-9}m$; iii) the characteristic frequency of the collective oscillations for a biomolecule
$\omega_{0}=10^{12}s^{-1}$ .
Moreover, since we are interested in observing self-emergent synchronization,
we consider a set of identical molecules in order to maximise the probability of observing it; 
therefore we assume $\widetilde{\mathcal{R}}_{i}=1$, $\widetilde{m}_{i}=10$  and $x_{i0}\simeq 5$ for all $i=1,\ldots N$ according to characteristic dimension and masses of biomolecules.\\
The parameter that fixes the characteristic length for the emergence of non linear phenomenon has been settled to be $\widetilde{\Lambda}\simeq 0.85$. 
The temperature of the system has been settled at $T=300 K$ and consequently for our choices $\mathcal{E}_{\mathrm{bath}}=2.5 \times 10^{-3}$,
while water viscosity is $\eta_{W}\simeq 8.54 \times 10^{-4}\, \mathrm{Pa\cdot s}$ and $\widetilde{\eta}_W=0.56$ yielding to $\Omega_{\mathrm{frict},i}=\Omega_{\mathrm{frict}}=1.05$.
With our choice of free parameters of the system,
the strength of thermal noise results $\widetilde{\Psi}\simeq 4.6 \times 10^{-2}$.

The frequencies associated to the electrodynamic interactions  $\Omega_{ij}^2$ can be expressed in terms of adimensionalized units
\begin{equation}
\Omega^{2}_{ij}=\dfrac{1}{\omega_0^2}\dfrac{2 e^2}{4\pi\epsilon_{WS}\mu\lambda^3}\dfrac{Z^2}{\widetilde{m}\widetilde{R}_{ij}^3}
\end{equation}
where $\widetilde{R}_{ij}$ is the mutual distance among the centers of the dipoles expressed
in unit of $\lambda$ and $\widetilde{m}$ is the mass of a molecule expressed in
adimensionalized units.
In the performed simulations the position of each dipole representing a biomolecule
is assigned in a cube box of unitary side, i.e. the components of the vector 
position of the center of each dipole have coordinates $\widetilde{\mathbf{R}}_{i}=(x_i,y_i,z_i)$,
with $x_i,y_i,z_i\in[0,N^{1/3}\langle \widetilde{d}\rangle]$, where $N$ is the total number of
dipoles and $\langle \widetilde{d}\rangle$ is the average intermolecular distance in $\lambda$
units.
As a reference case in our simulations the parameters have been chosen to be
$\widetilde{m}=10$, $Z_{i}=1000$, while the average intermolecular
distance $\langle \widetilde{d}\rangle=\lambda \langle \widetilde{d}\rangle=1.6\times 10^{3} \AA=1.6\times 10^{-7}m$.
The reason for choosing such a large value of $Z$ is justified under the hypothesis that the surrounding
water molecules participate to the effective dipole of each biomolecule and enhance it.
Therefore for the considered choice of parameters $\Omega_{ij}^2\sim 2.3\times 10^{-3} $. 
Finally, in order to consider different cases with stronger interactions (corresponding to shorter average intermolecular
distances, for instance) the coupling term is multiplied by a factor $K>1$ with respect
to the reference case just discussed.

The  parameter  $\mathcal{E}_{\mathrm{pul}}$ can be estimated assuming that the energy
injection on each biomolecule is due to the vibrational decay of a fluorescent dye.
It is realistic \cite{NardecchiaTorres} to consider a difference between the absorbed and emitted frequency
of the order of $\Delta \nu_{\mathrm{fluor}}\simeq 5\times 10^{13}\mathrm{s}^{-1}$ 
and $\widetilde{m}_{\mathrm{fluor}}\simeq 0.6$  yielding to $\mathcal{A}_{NE}\simeq 1.4\times 10^{-2}$. 
The characteristic frequency for the energy transfer $\Omega_{\mathrm{pul}}$  is one of the most delicate
parameters to be settled. As this term in principle accounts for the continuous injection of energy into the system,
but the release must be done without perturbing too much the oscillating behavior, we can assume that  $\Omega_{i}\gg \Omega_{\mathrm{pul}}\simeq 10^{-2}$.

\section{Numerical Results}

\begin{figure}[h!]
\begin{center}
\includegraphics*[angle=0,width=8.3cm]{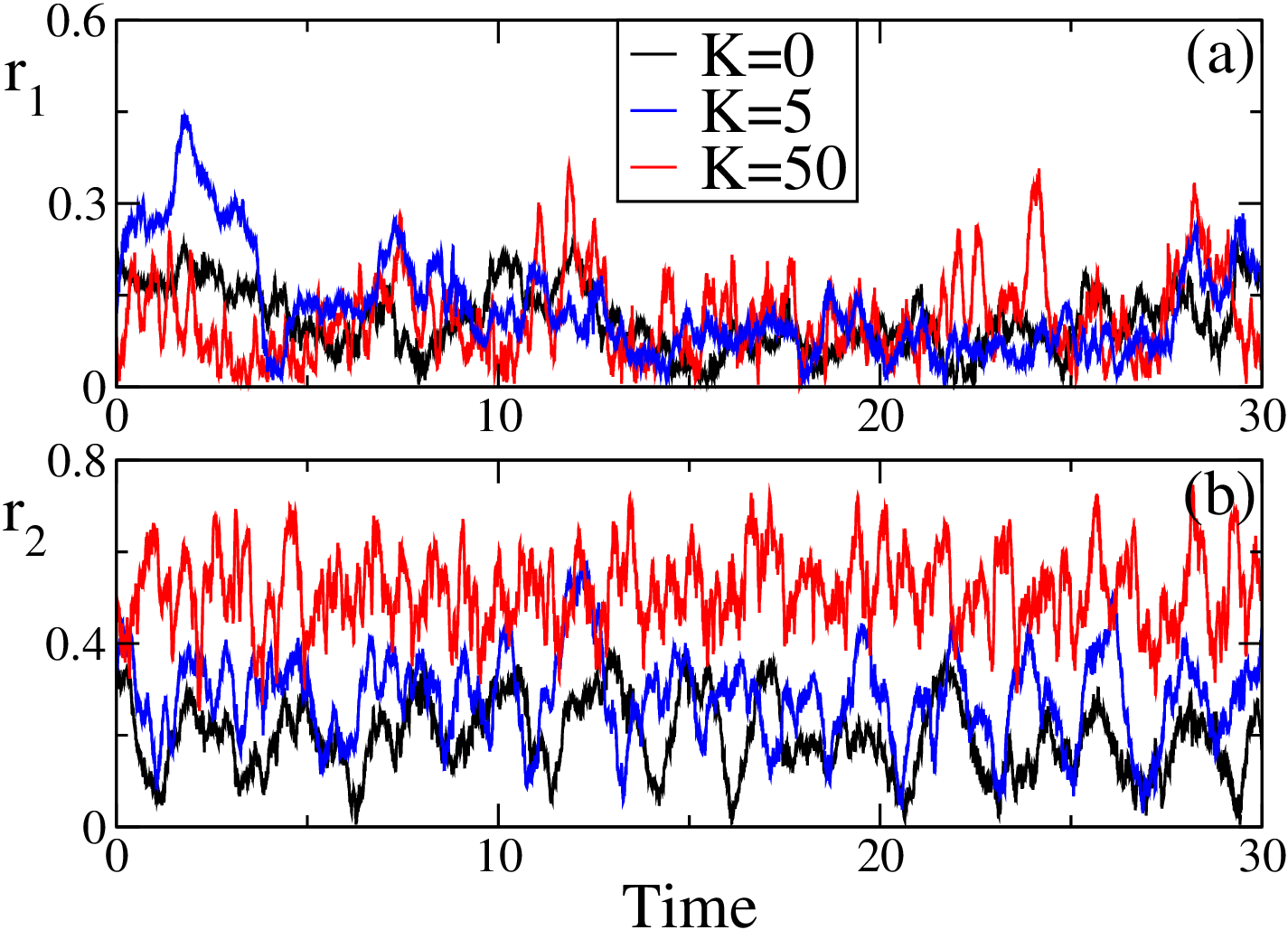}
\includegraphics*[angle=0,width=8.cm]{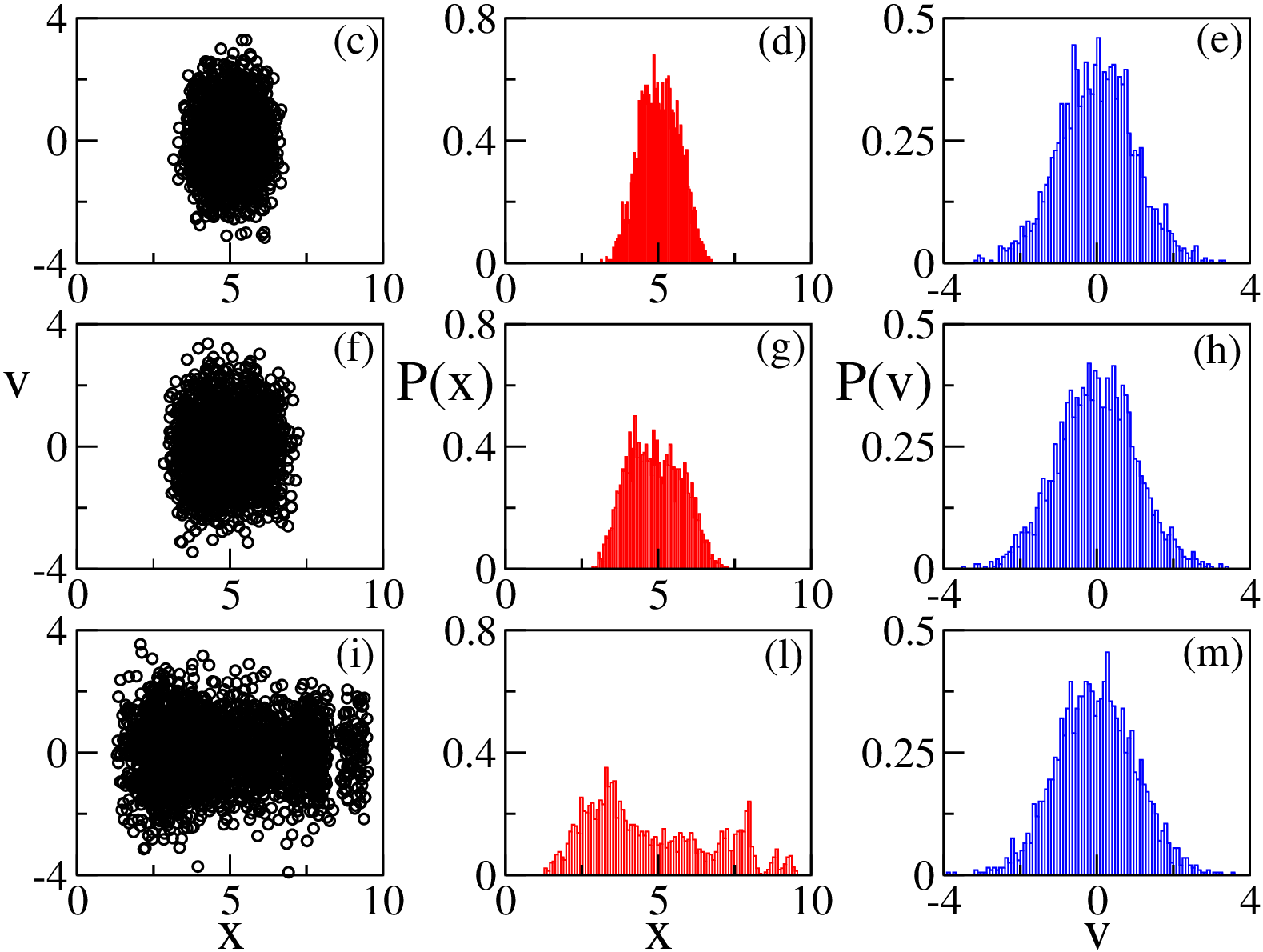}
\end{center}
\caption{ Synchronization properties of the system. Order parameters $r_1$ (a), $r_2$ (b) as a function of time for different coupling constants. 
Panels (c),(f), (i): snapshots of the velocities of the single dipoles as a function of their positions for K=0 (c), K=5 (f), K=50 (i). 
Panels (d), (g), (l): probability distribution of the positions of the dipoles for different coupling constants. The panels refer 
to K=0 (d), K=5 (g), K=50 (l). Panels (e), (h), (m): probability distribution of the velocities of the dipoles for different coupling constants. The panels refer 
to K=0 (e), K=5 (h), K=50 (m). The parameters values used for these simulations are: $\Omega_i=1$, $x_{i0}=5$, $\Omega_{frict,i}=0.105$ (for every $i=1,\ldots,N$),
$\Omega_{pul}=0.1$, $\mathcal{A}_{NE}=1.4$.
}
\label{fig.0}
\end{figure}

The reported analyses have been done using a single system size (N=50) and random initial conditions both for positions and velocities. 
However, similar results have been obtained for N=100, 200 (not shown). The collective evolution of the population and in particular the level of coherence is
usually characterized in terms of the macroscopic field
\begin{equation}
 \rho(t)= r_1(t)e^{i\Phi(t)}=\frac{1}{N}\sum_{j=1}^N e^{i\theta_j(t)},
\end{equation}
where the modulus $r_1$ is an order parameter for the synchronization transition being one ($\mathcal{O}(N^{-1/2})$) for synchronous (asynchronous) states, while
$\Phi$ is the phase of the macroscopic indicator \cite{Winfree}. However, in our case, the molecules are pivoted to the center of mass and cannot rotate: the effective degree of freedom
of these objects consists in an elongation/shrinkage along the direction identified by the mutual distance between the two centers of charges. Therefore it is not possible to 
describe the movement of the dipole in terms of an oscillator rotating along the unit-circle via the identification of a time-dependent phase. The solution that we have adopted is to calculate
the phase of the single molecule by using the inversion formulas
\begin{equation}
 \sin\theta_i=\frac{x_i-x_{0i}}{\sqrt{v_i^2 +(x_i-x_{0i})^2}}, \qquad \cos\theta_i= \frac{v_i}{\sqrt{v_i^2 +(x_i-x_{0i})^2}}
\end{equation}
to associate a phase $\theta_i \in [-\pi, \pi]$ according to
\begin{eqnarray}
 \theta_i =
  \begin{cases}
    \arcsin(\sin\theta_i)     & \quad \text{if } \cos\theta_i\geq 0\\
    \pi -\arcsin(\sin\theta_i)     & \quad \text{if } \sin\theta_i>0 \wedge \cos\theta_i< 0\\
    -\pi -\arcsin(\sin\theta_i)     & \quad \text{if } \sin\theta_i<0 \wedge \cos\theta_i< 0.\\
  \end{cases}
\end{eqnarray}
However the calculation of the order parameter $r_1$ does not lead to the identification of emergent (phase) synchronization in the system; in particular $r_1$ does not show any dependence on the 
coupling constant (see Fig. \ref{fig.0}(a)), as we would expect when the molecules are interacting with increasing strength. In addition to this, the emergence of a collective behavior is not identifiable 
in a straightforward manner neither looking at the order parameter usually employed to identify the emergence of 2-clusters ($r_2(t)=| \frac{1}{N}\sum_{j=1}^N e^{i 2\theta_j(t)}|$), nor at the distribution of 
positions and velocities of the molecules (see Fig. \ref{fig.0}, panels (b)-(m)). Looking at the phase space $(x,v)$ it does not emerge a clear separation in synchronized clusters among the dipoles and 
also the probability distributions of positions and velocities are simply Boltzmann-distributed, as we expect from a set of indepent oscillators subjected to a single asymmetric well potential in absence 
of coupling. Only for very strong coupling (K=50) we can observe the emergence of a secondary small cluster in the phase space $(x,v)$ (see Fig. \ref{fig.0}(i)) that leads to a modification of the 
probability distribution of the positions, that is no more simply Boltzmann-distributed, and to an increasing of the average value of $r_2(t)$ (see Fig. \ref{fig.0} panels (l) and (b) respectively).

Therefore, in order to investigate the emergence of a collective behavior due to the interactions among the molecules we consider the variable 
\begin{equation}
\label{eq:PolarizationMod}
 P(t)=\sqrt{\sum_{i=1}^N \left\lbrace [(x_i(t)-x_{i0})\sin\beta_i \cos\phi_i]^2 + [(x_i(t)-x_{i0})\sin\beta_i \sin\phi_i]^2 +[(x_i(t)-x_{i0})\cos\beta_i]^2\right\rbrace }
\end{equation}
which represents the ensemble average of the projection of the dipole position in the cartesian coordinates system $X, Y, Z$. The biomolecule in our model is identified
via the intermolecular mutual distance between the two centers of charges measured along the radial $\textbf{x}$ direction and we need to express this variable in cartesian coordinates.  
In other words, each term under the square root represents the component of the dipole position along one of the directions $X, Y, Z$, thanks to the respective projection angle
$\beta_i$ of each molecule's radius to the Z-axis and $\phi_i$ of the projection of $x_i$ in the XY plane to the X-axis. These angles are generated together with the initial conditions and do not vary in time. 

Due to the fact that the system is not deterministic and a white noise source is present into the differential equations, we have developed
a method similar to the second-order Runge-Kutta one for solving numerically ordinary differential equations. In particular we have implemented
the Heun method \cite{Heun} in the Runge-Kutta algorithm as suggested in \cite{Toral}, and we have used an integration time step 0.002 to perform the simulations.
In addition to this, in order to compare the results for different coupling constant values and for different strengths of the thermal noise, 
we implemented a low-pass filter to analyse the power spectra. This filter relies on the differentiation properties of the Fourier transform;
in particular, since the Fourier transform of a generic function $f$ is related to the Fourier transform of its derivative via the relationship 
$\mathcal{F}\left[\frac{\partial f(\textbf{x})}{\partial x_j}\right] = 2\pi i \nu_{j}\widehat{f}(\nu)$, it is possible to filter the low-frequency components 
of the spectrum just using the Fourier transform of the derivative.

\begin{figure}[h!]
\begin{center}
\includegraphics*[angle=0,width=12.cm]{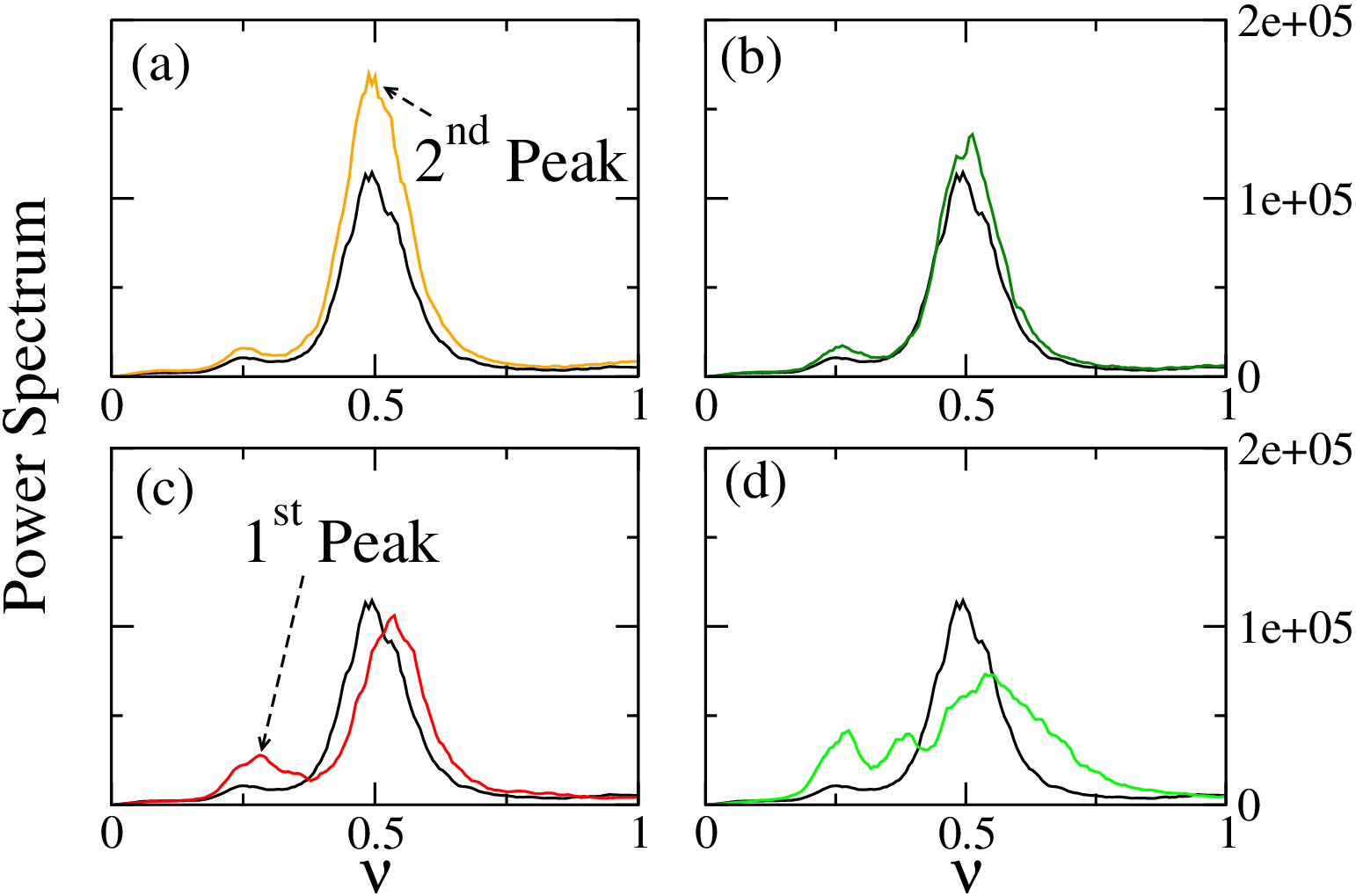}
\includegraphics*[angle=0,width=12.cm]{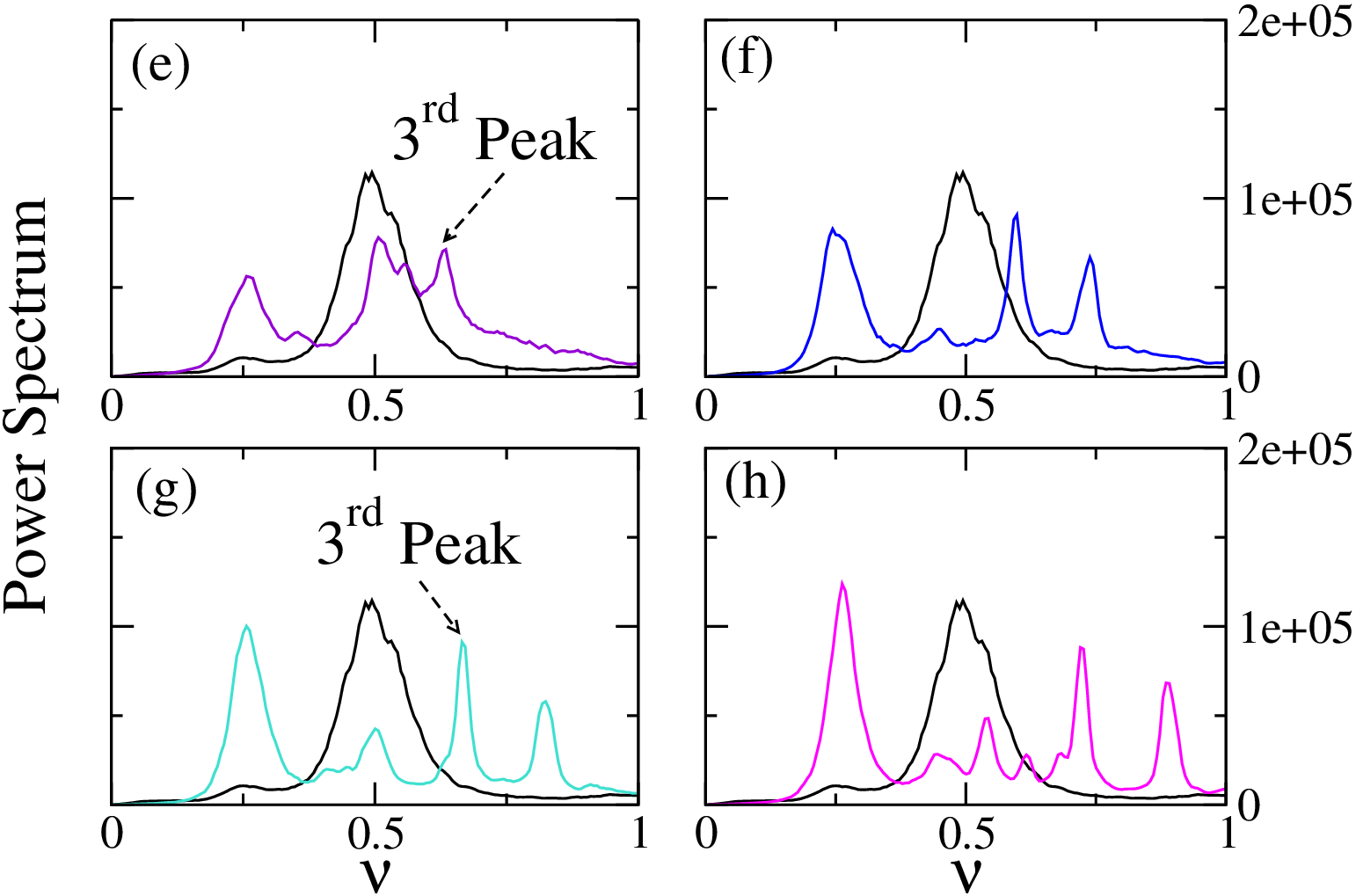}
\end{center}
\caption{Investigation of the emergence of a collective behavior as a characteristic peak in the power spectrum. Panels (a)-(h): Power spectrum of $dP/d\textbf{x}$ for different values of the 
coupling constant K and for thermal noise strength $\tilde{\Psi}_i=0.46$.
The black curve represents, in each panel, the power spectrum of the system without coupling (K=0). The other curves shown are, respectively, for $K=1$ (a); $K=2$ (b);
$K=5$ (c), $K=10$ (d); $K=21$ (e); $K=31$ (f); $K=41$ (g); $K=50$ (h). Other parameters as in Fig. \ref{fig.0}.
}
\label{fig.1a}
\end{figure}

\begin{figure}[h!]
\begin{center}
\includegraphics*[angle=0,width=7.5cm]{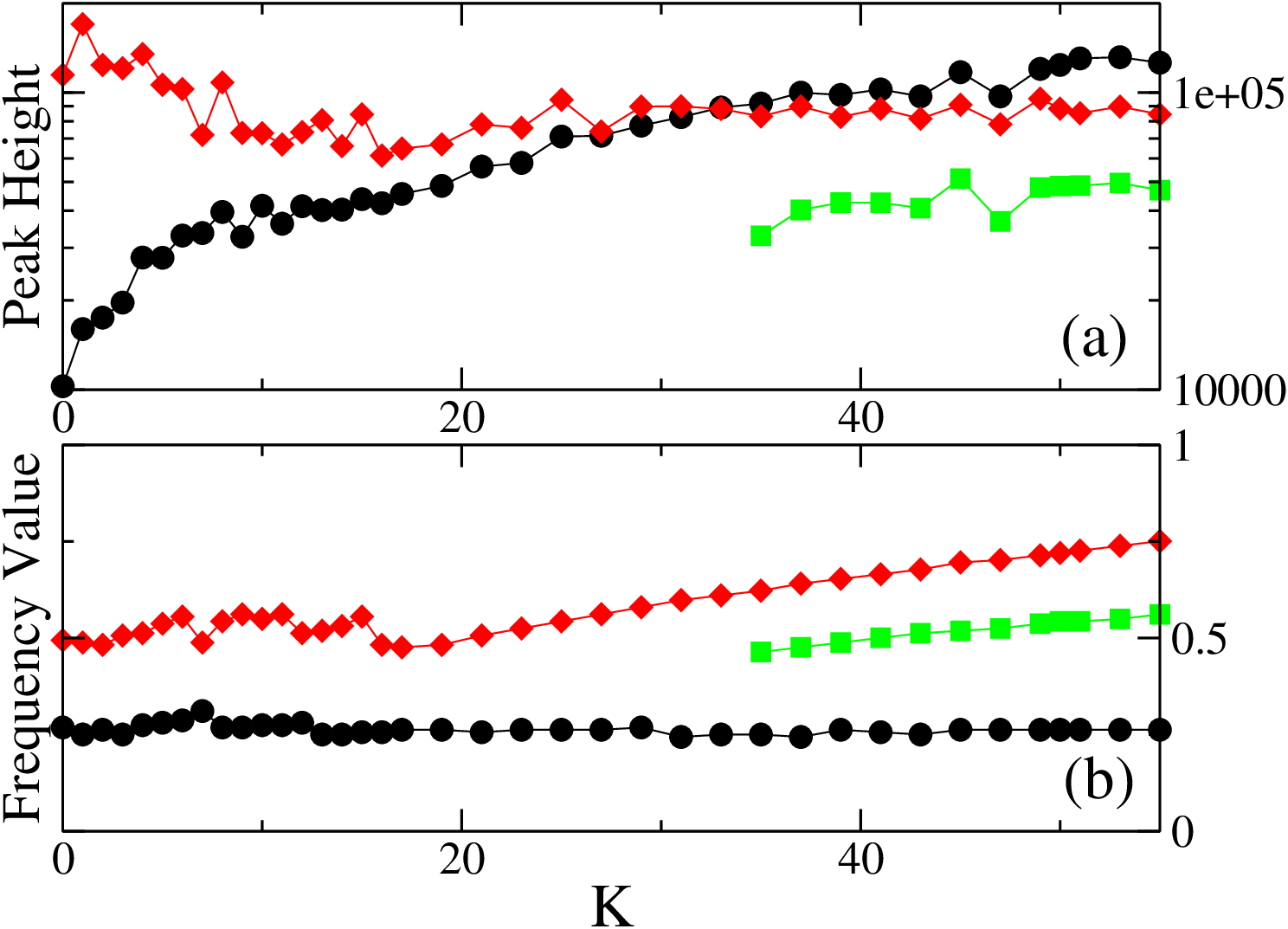}
\includegraphics*[angle=0,width=7.7cm]{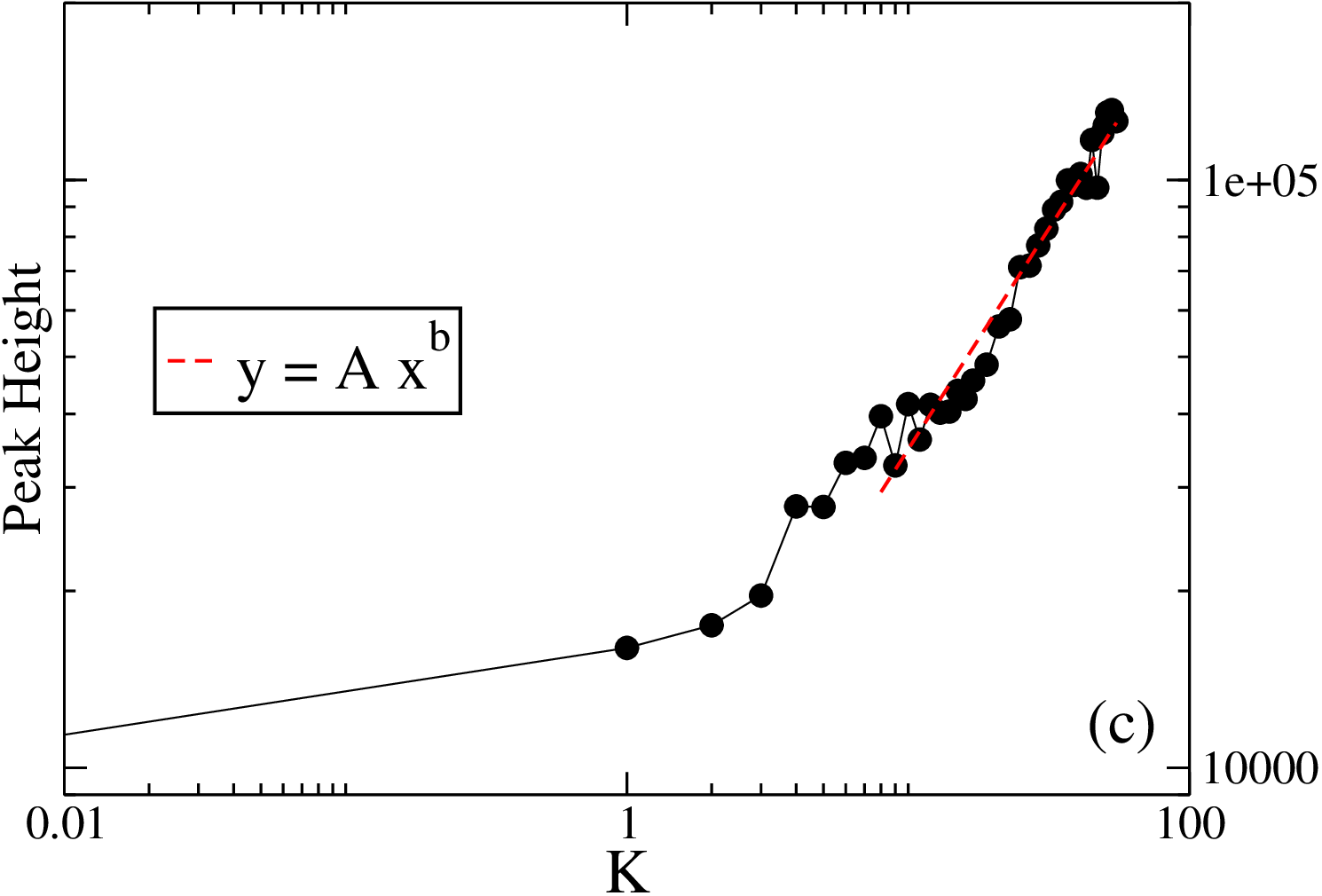}
\end{center}
\caption{Dependence of the system's characteristic frequencies on the coupling constant. Panels (a), (b): Peak height (a) and frequency value (b) of the first three main peaks that 
characterize the dynamics of the system. Panel (c): Fitting of the dependence of the peak height on the coupling constant. Fitting values are $A=6188,4\pm 0.5$, $b=0.75\pm 0.03$. 
For all the panels the black dotted curve represents the first peak, the red diamonds curve represents the second peak and the square green curve represents the third peak.
Parameters as in Fig. \ref{fig.0}.
}
\label{fig.1b}
\end{figure}

Therefore we calculated the power spectrum of $dP/d\textbf{x}$ to investigate the role played by the interactions among the dipoles to enhance a collective motion.
While in absence of interactions (K=0), the system shows a single pronounced peak at frequency  $\approx 0.488\pm 0.006$, once the interactions are active $(K>0)$, another peak
arises at smaller frequency $\approx 0.263\pm 0.013$. By increasing the value of K we observe an increase of the peak at lower frequency, to which corresponds a decreasing of the peak
at higher frequency: a collective motion is enhanced due to interaction, while the motion corresponding to the non-connected situation is depressed (see Fig. \ref{fig.1a},
panels (a)-(h) and Fig. \ref{fig.1b}(a)). On the other hand the position of the peak (i.e. the corresponding frequency value) does not change significantly if we increase the coupling
constant (see Fig. \ref{fig.1b}(b)); the more evident increasing ratio for $K>20$ is related to the fact that power spectra become richer and richer for higher coupling and 
secondary peaks arise. One of these secondary peaks (the main one) emerging at bigger coupling constant is also reported in Fig. \ref{fig.1b} (panels (a), (b)), and it is termed ``Third Peak''.

Finally, if we analyse in more details the behavior of the first peak, related to the emergent collective motion, as a function of the coupling contant, it is possible to identify two different
scales, once the figure is plotted in log-log scale (Fig. \ref{fig.1b}(c)). In particular, the different scales present for low coupling constant ($K<5$) and for sufficiently strong coupling ($K>10$)
denote a transition between two different dynamical behaviors: the cross-over between two different regimes, from the one dominated by individual asynchronous behavior, to the one dominated by collective motion,
with strongly interacting oscillators, is thus compatible with these two different scales.

\begin{figure}[h!]
\begin{center}
\includegraphics*[angle=0,width=8.4cm]{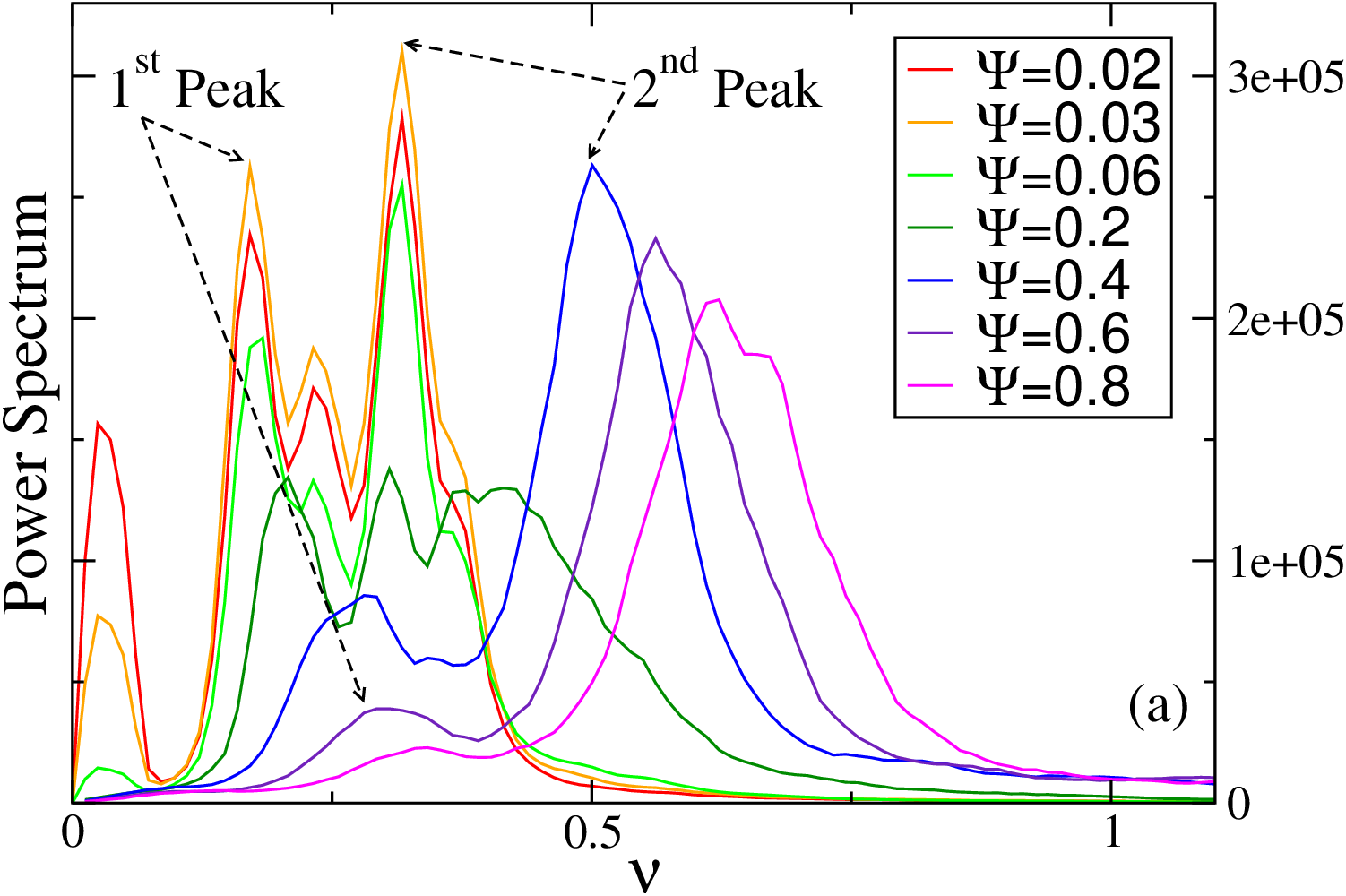}
\includegraphics*[angle=0,width=7.9cm]{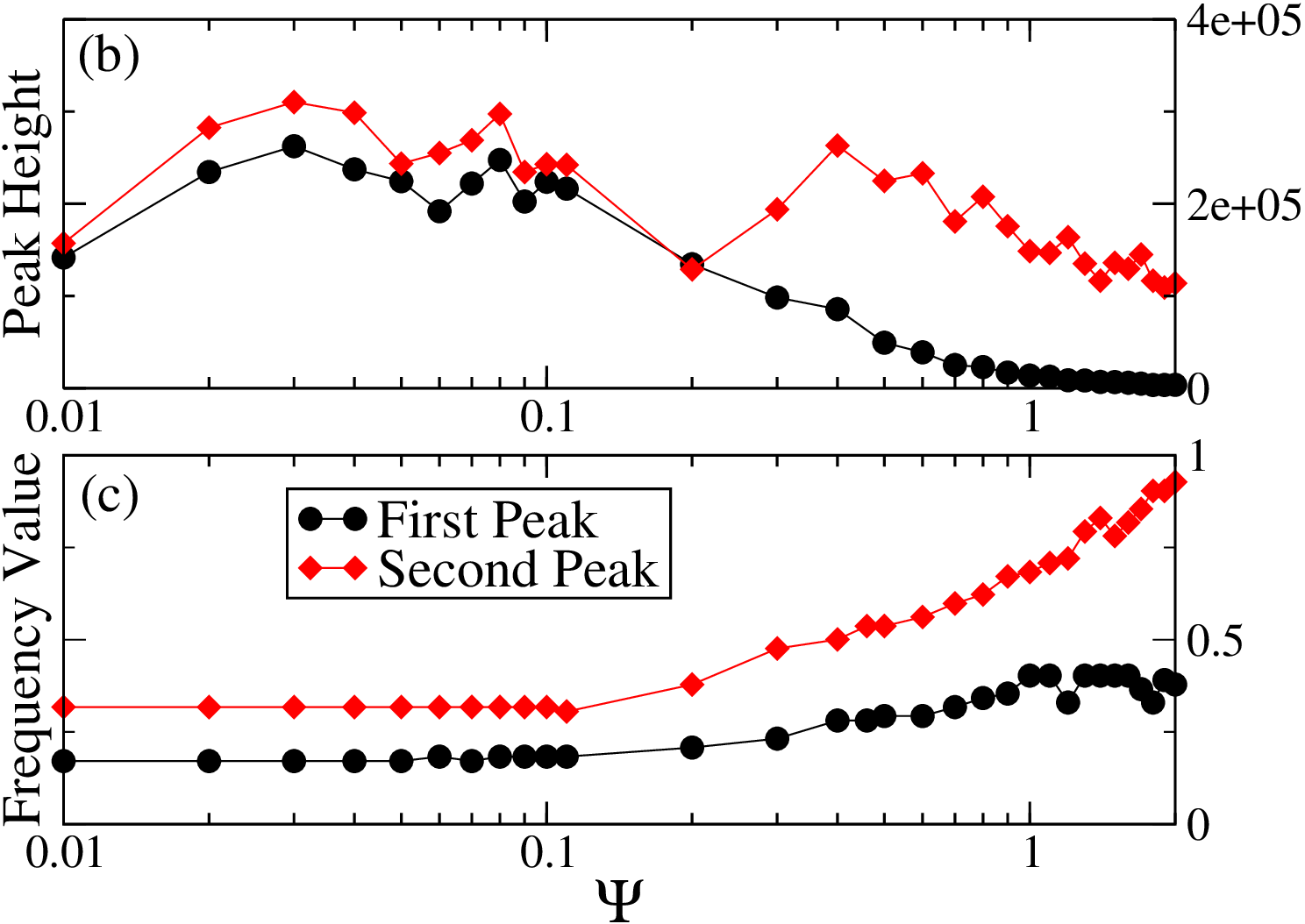}
\end{center}
\caption{Response of the system under the effect of the thermal noise strength. Panel (a): Power spectrum of $dP/d\textbf{x}$ for different values of the thermal noise strength and for coupling constant K=5.
Panels (b), (c): Peak height (b) and frequency value (c) of the first two main peaks that characterize the dynamics of the system.
Parameters as in Fig. \ref{fig.0}. The values of the different thermal noise strengths reported in the caption of panel (a) and the axix label
in panel (c) must be intended as $\tilde{\Psi}$: the $\enskip\tilde{}\enskip$ has been suppressed in the figure for the sake of simplicity.
}
\label{fig.2}
\end{figure}

If we now investigate the response of the system under the effect of the thermal noise strength, we obtain a stochastic resonance effect \cite{Gammaitoni}:
the signal at low frequency ($\approx 0.28\pm 0.09$) can be boosted by adding white noise to the signal, which contains a wide spectrum of frequencies. 
The frequencies in the white noise spectrum corresponding to the original signal's frequencies resonate with each other, thus amplifying the original signal (i.e. the signal at
low frequency) while not amplifying the rest of the white noise. Furthermore the signal-to-noise ratio is increased, while 
the added white noise is filtered out thanks to the band-pass filter that we have implemented calculating the power spectrum of $dP/d\textbf{x}$.
In particular the low frequency peak, that corresponds in our case to the collective motion, is more visible for thermal noise strength $\tilde{\Psi}=0.03$, to which 
corresponds a maximum in the peak high (see Fig.\ref{fig.2} panels (a),(b)). This peak is depressed for higher temperature and less likely to be revealed.
On the other hand the peak at high frequency ($\approx 0.56 \pm 0.22$), corresponding to the dynamics of isolated dipoles, 
can be also boosted by adding white noise into the system, but it does not decrease as significantly as the former one for higher temperatures,
thus meaning that the single dipoles in this model are able to react to big level of noise, even though this is physically not plausible, since we would expect
that dipoles will break up for high temperatures.

\section{discussion}
Let us now comment about the physical meaning, and about the prospective relevance, 
of the results described in the previous Sections.
The present work was motivated by the need of finding an experimental strategy – complementary 
to the diffusion based one already suggested in \cite{pre1,pre2,Nardecchia2017} - to detect an intermolecular long range 
electrodynamic interactions, if any.
The background scientific framework is the following. By pumping energy in the biomolecules of a 
watery solution, that is by keeping these molecules warmer than the solvent (out-of-thermal equilibrium), 
when the input energy rate exceeds a threshold value, then all, or almost all, the excess energy 
(that is,  energy input minus energy losses due to dissipation) is channeled into the vibrational mode 
of the lowest frequency.  In other words, the shape of the entire molecule is periodically deformed 
resulting in a ``breathing'' movement \cite{NardecchiaTorres}. In so doing the biomolecules behave as microscopic 
antennas that absorb the electromagnetic radiation tuned at their ``breathing'' (collective) oscillation frequency. But antennas at the same time 
absorb and re-emit electromagnetic radiation, thus, according to a theoretical prediction, these antennas 
(biomolecules) can attractively interact at a large distance through their oscillating near-fields, and through 
the emitted electromagnetic radiation, if these oscillations are resonant, that is, take place at the same frequency \cite{NardecchiaTorres}.
The still open question is whether these electrodynamic interactions can be strong enough to be experimentally 
detectable, and ultimately relevant to biology. In our schematic modeling of a watery solution of biomolecules,
we have then assumed that the above mentioned collective vibrations of each individual molecule are present so 
that they interact with a potential falling as $1/r^3$ with the intermolecular distance $r$. By adopting physically 
reasonable values for the molecular parameters entering the equations of motion of the molecular dipoles, we have 
numerically investigated the effect of varying the mutual dipole-dipole electrodynamic interactions by changing the 
parameter $K$. The novel phenomenon observed and reported in the preceding Section is the 
appearance of a spectral signature of an intermolecular collective phenomenon which manifests itself with an 
increasing evidence when the parameter $K$ is raised. Physically, this suggests that the stepping up of supposedly 
activated electrodynamic intermolecular interactions could be, in principle, spectroscopically detected by varying 
the concentration of the soluted biomolecules. This latter fact, of course, entails the variation of the average interparticle 
distance $\langle d\rangle$ according to the relation $\langle d\rangle =C^{-1/3}$, where $C$ is the concentration 
of the solution. And varying $C$ would be a practical way of experimentally varying the parameter $K$.
In order to detect the emergence of a collective behavior due to the interactions among the molecules we considered 
the variable $P(t)$ in Eq.\eqref{eq:PolarizationMod} representing the ensemble average of the projection of the dipole positions in the 
cartesian coordinates system. Strictly speaking, this is not yet directly spectroscopically measurable, but it is tightly 
related with the overall dipole moment of the solution that could be more directly spectroscopically accessible. 
However, this is a technical detail which will be more thoroughly addressed while designing a specific experiment. 
For the moment being, the results reported in the present work outline a very promising strategy - complementary 
to the diffusion based one - to reach a proof of concept, or a refutation, of the possible relevance of long range 
electrodynamic intermolecular interactions to our understanding of the biochemical machinery at work in living matter.

\appendix
\section{Discussion about characteristic time scales on the system}\label{sec:TimScales}
The characteristic frequency for giant dipole oscillations has been conjectured to lie in a range between $0.1-10$ THz so that the characteristic time for the oscillations is $\tau_{osc}\sim\omega_{0}^{-1}\simeq 10^{-13}-10^{-11} s$.
Recent experiments seem to provide a first evidence of the existence of collective biomolecule oscillations in this range of frequency in out-of-thermal equilibrium conditions.
The characteristic time scale associated with translational diffusion of biomolecules can be estimated by
\begin{equation}
\label{eq:tsc_trs}
\tau_{trs}\approx\dfrac{\delta R^2}{6D_{trs}}
\end{equation}
where $\delta R$ is the tolerance in defining center of mass position 
of two biomolecules and  $D_{trs}$ is the self-diffusion coefficient of a biomolecule. We are interested in studing collective phenomena emerging due to long-range
interactions in ``diluted" system, meaning that the average
intermolecular distance $\langle R \rangle\approx 10^{-7}\mathrm{m}$ is much larger then the characteristic molecular linear dimension scale $\lambda_{bio}\gtrsim10^{-9}\mathrm{m}$ 
of biomolecules: this allows to consider $\delta R\approx \lambda_{bio}$. Using Einstein's formula for Brownian self-diffusion coefficient $D_{trs}=k_{B}T/(6\pi\eta_{W}\lambda_{bio})$ 
in eq. \eqref{eq:tsc_trs} we obtain
\begin{equation}
\tau_{trs}\simeq\dfrac{6\pi\lambda_{bio}^3}{k_{B}T}\gtrsim 3 \times 10^{-9} s\approx 10^{2} \tau_{osc}.
\end{equation}
This makes plausible the hypothesis that the center of mass of each biomolecule can be considered as a parameter and not a dynamical variable.
Analogously, the characteristic time for biomolecules rotational diffusion
has been estimated using
\begin{equation}
\tau_{rot}\approx D_{rot}^{-1}=\left(\dfrac{k_{Bol}T}{8\pi\eta_{W}\lambda_{bio}^3}\right)^{-1}\gtrsim 5 \times 10^{-9} s \approx 10^{2} \tau_{osc}\,.
\end{equation} 
It follows that also diffusive rotation can be neglected on time scales characteristics for giant dipole oscillations and the orientation of
dipole can be assumed to be initially fixed.

%The system, prepared with random initial conditions, behaves, in absence of coupling, like a set of indepent oscillators subjected to a single asymmetric well potential.
%Therefore we expect no indication of synchronization and Boltzmann-distributed probability functions for positions and velocities.

\newpage
\begin{acknowledgments}
The authors acknowledge the financial support of the Future and Emerging Technologies (FET) Program within the Seventh Framework Program (FP7) for Research of the European Commission, 
under the FET-Proactive TOPDRIM Grant No. FP7-ICT-318121. S. O. thanks Stefano Lepri for useful discussions and suggestions and she
acknowledges the Deutsche Forschungsgemeinschaft via Project A1 in the framework of SFB 910.
\end{acknowledgments}

\section*{Author Contributions}
S.O. performed the numerical simulations. M. G. elaborated the dynamical model.
M. G. and S. O. prepared the manuscript. All the authors developed the theoretical methods and reviewed the manuscript. As team leader, M.P.
supervised all the aspects of the work.

\section*{Additional Information}
Competing financial interests: The authors declare no competing financial interests.

\end{document}